\documentclass{aa}
\usepackage[varg]{txfonts} 

\usepackage{graphicx} 

\usepackage{amsmath} 

\begin{document}
\title{On the cosmic previrialization \thanks{To be submitted to A\&A}}


\author{
Shubhadip Bera,\inst{1,2}\thanks{E-mail: 
sbera@camk.edu.pl}
\and
Micha{\l} J.\ Chodorowski\inst{2}
}

\institute{
Nicolaus Copernicus Superior School (SGMK), Nowogrodzka 47a, 00-695 Warsaw, Poland
\and
Nicolaus Copernicus Astronomical Center, Bartycka 18, 00--716 Warsaw,\,Poland
}

\date{}
\abstract
{Non-linear gravitational evolution affects the growth of density and velocity fluctuations and leads to previrialization effects before virialized structures are formed. Therefore, understanding and quantifying the leading nonlinear corrections to the corresponding variances is crucial for describing the departure from linear structure formation.}
{We calculate the first nonlinear correction to the variance of the cosmic density and velocity fields.}
{We analyze the 1-loop contributions associated with the second and third order perturbative kernels of both the density and velocity-divergence fields. The relevant form of the third-order kernels $F_3(\mathbf{k},\mathbf{q},-\mathbf{q})$ for density, and $G_3(\mathbf{k},\mathbf{q},-\mathbf{q})$ for velocity divergence, are clearly non-symmetric under exchange of $\mathbf{k}$ with $\mathbf{q}$. However, the calculation of the variance involves a symmetric double integral of these variables, so in this {\em particular} case the kernels can be additionally symmetrized. We therefore explicitly symmetrize these kernels $F_3$ and $G_3$, and express them as functions of the two scalar variables: the cosine of the angle between the two wavevectors and a symmetric function of their magnitudes. This representation enables us to identify and isolate the terms which are divergent {\em before} performing the angular integration, and to show their cancellation between the second- and third-order terms. }
{The divergent contribution from $P_{\delta 22}$ to the one-loop power spectrum cancels exactly that arising from $P_{\delta 13}$. Angularly averaged divergence-free part of $F_3$ turns out to be remarkably close to a constant.}
{The coefficient of the nonlinear correction to the linear variance of the density field is in a narrow range, from $4007/2205$ to $4063/2205$ (approximately $1.817$ to $1.843$), for {\em any} form of the linear power spectrum.}


\maketitle
\nolinenumbers
\section{Introduction}
\hspace{.5cm}Peebles' conjecture of previrialization -- the early suppression of density fluctuations prior to virialization -- was formulated long ago \citep{Davis}, yet its quantitative understanding remains incomplete. Although cosmological perturbation theory can address this phenomenon within a limited regime, it gives some valuable insight into the beginning of nonlinear effects. 

The leading nonlinear corrections to the linear-theory variances of the cosmic density and velocity fields were calculated by \citet{Scocci} using a diagrammatic approach to Eulerian perturbation theory. Here in this work, we present a more direct and transparent alternative method based on both the exact dynamics and the Zel'dovich Approximation for deriving the leading-order corrections readily extendable to higher-order power-law power spectra.
\label{sec:intro}
\section{Calculations}
The power spectrum of cosmic density fluctuations up to the leading-order non-linear corrections reads
\begin{equation}
P_{\rm 1Loop}(k)=P_{11}(k)+P_{22}(k)+P_{13}(k),
\label{eq:p1loop}
\end{equation}
where $P_{11}(k)\equiv P(k)$ is the power spectrum calculated in the framework of linear theory,
\begin{equation}
P_{22}(k)=\frac{2}{(2\pi)^3} \int \left[F_2(\mathbf{k}-\mathbf{q},\mathbf{q})\right]^2 P(|\mathbf{k}-\mathbf{q}|)P(q)\,\mathrm{d}^3q,
\label{eq:p22}
\end{equation}
and
\begin{equation}
P_{13}(k)=\frac{6}{(2\pi)^3}P(k)\int F_3(\mathbf{k},\mathbf{q},-\mathbf{q})P(q)\,\mathrm{d}^3q.
\label{eq:p13}
\end{equation}
The quantities $F_2$ and $F_3$ are respectively the second and the third-order symmetrized perturbative kernels of the density field which are homogeneous functions of order zero. Their explicit forms will be given in the following subsections. The one-loop density variance reads therefore
\begin{equation}
\sigma^2=\sigma_L^2+\sigma_{22}^2+\sigma_{13}^2.
\label{eq:sigma-total}
\end{equation}
Here, $\sigma_L^2\equiv\sigma_{11}^2$ is the variance in the linear approximation,
\begin{equation}
\sigma_{22}^2=\frac{2}{(2\pi)^6}\int \mathrm{d}^3k\,\mathrm{d}^3q\,P(k)P(q)[F_2(\mathbf{k},\mathbf{q})]^2
\label{eq:sigma22}
\end{equation}
and
\begin{equation}
\sigma_{13}^2=\frac{6}{(2\pi)^6}\int \mathrm{d}^3k\,\mathrm{d}^3q\,P(k)P(q)F_3(\mathbf{k},\mathbf{q},-\mathbf{q}).
\label{eq:sigmaNonSym13}
\end{equation}
As both $\mathbf{k}$ and $\mathbf{q}$ are dummy variables of integration here, the latter integral will be identical when we replace $F_3(\mathbf{k},\mathbf{q},-\mathbf{q})$ with $F_3(\mathbf{q},\mathbf{k},-\mathbf{k})$. We can thus write
\begin{equation}
\sigma_{13}^2=\frac{6}{(2\pi)^6}\int \mathrm{d}^3k\,\mathrm{d}^3q\,P(k)P(q)F_{3s}(\mathbf{k},\mathbf{q}),
\label{eq:sigma13}
\end{equation}
where
\begin{equation}
F_{3s}(\mathbf{k},\mathbf{q})\equiv\frac{1}{2}\left[F_3(\mathbf{k},\mathbf{q},-\mathbf{q})+F_3(\mathbf{q},\mathbf{k},-\mathbf{k})\right].
\label{eq:F3Sym}
\end{equation}

\subsection{Calculation of $\Delta\sigma_{\delta22}^2$ }
\label{sec:sigma22}

The kernel $F_2$ can be written in the following way
\begin{equation}
F_2(\mathbf{k},\mathbf{q})=c_0+a^{-1}\mu_{kq}+c_2\,\mu_{kq}^2\,,
\label{eq:F_2}
\end{equation}
where $c_0+c_2=1$; specifically
\begin{equation}
c_0=5/7,\quad c_2=2/7\quad ({\rm ED});\qquad c_0=c_2=1/2\quad ({\rm ZA}).
\label{eq:c0c2}
\end{equation}
The acronyms `ED' and `ZA' stand here respectively for Exact Dynamics and the Zel'dovich Approximation. The quantity $\mu_{kq}$ is the cosine of the angle between vectors $\mathbf{k}$ and $\mathbf{q}$, $\mu_{kq}=\mathbf{k}\cdot\mathbf{q}/(kq)$. The quantity $a$ is
\begin{equation}
a(k,q)=\frac{2kq}{k^2+q^2}\,.
\label{eq:a}
\end{equation}
Clearly, $a\in[0,1]$ and $a^{-1}$ diverges when either $k/q$ or $q/k$ converge to zero. We decompose $2F_2^2$ in the following way:
\begin{align}
2[F_2(\mathbf{k},\mathbf{q})]^2={}&b_0L_0(\mu_{kq})+b_1L_1(\mu_{kq})+b_2L_2(\mu_{kq}) , \nonumber \\
&+b_3L_3(\mu_{kq})+b_4L_4(\mu_{kq})+2a^{-2}\mu_{kq}^2 \,,
\label{eq:F2Non}
\end{align}
where $L_l(\mu_{kq})$ are the Legendre polynomials of order $l$. The reason for extracting the most divergent term will become evident soon. The coefficients $b_l$ are
\begin{align}
b_0&=2\left(c_0^2+\frac{2}{3}c_0c_2+\frac{1}{5}c_2^2\right), \nonumber\\
b_1&=4\left(c_0+\frac{3}{5}c_2\right)a^{-1}, \nonumber\\
b_2&=8\left(\frac{c_0}{3}+\frac{c_2}{7}\right)c_2, \nonumber\\
b_3&=\frac{8}{3}c_2a^{-1},\qquad b_4=\frac{16}{35}c_2^2.
\label{eq:b_j}
\end{align}
We have
\begin{equation}
\begin{split}
\sigma_{\delta22}^2={}&\int\frac{\mathrm{d}k\,\mathrm{d}q}{(2\pi^2)^2}k^2P(k)q^2P(q)
\sum_{l=0}^{4}b_l\int\frac{\mathrm{d}\Omega_k\,\mathrm{d}\Omega_q}{(4\pi)^2}L_l(\mu_{kq})\\
&+\frac{1}{(2\pi)^6}\int\mathrm{d}^3k\,P(k)\int\mathrm{d}^3q\,P(q)\frac{2\mu_{kq}^2}{\left[a(k,q)\right]^2} \,.
\label{eq:sigma22spher}
\end{split}
\end{equation}
Angular integration of the Legendre polynomial part is trivial, as $\int\mathrm{d}\Omega_1\mathrm{d}\Omega_2\,L_l(\mu_{12})=\int\mathrm{d}\Omega_1\mathrm{d}\Omega_2\,L_l(\mu_{12})L_0(\mu_{12})=(4\pi)^2\delta_{l0}$. We kept the integral of the last term in Eq.~(\ref{eq:F2Non}) intact. The reason for this will be evident soon. We finally obtain
\begin{equation}
\begin{split}
\sigma_{\delta 22}^2={}&2\left(c_0^2+\frac{2}{3}c_0c_2+\frac{1}{5}c_2^2\right)\sigma_L^4\\
&+\frac{1}{(2\pi)^6}\int\mathrm{d}^3k\,P(k)\int\mathrm{d}^3q\,P(q)\frac{2\mu_{kq}^2}{[a(k,q)]^2}.
\label{eq:sigma22DivNon}
\end{split}
\end{equation}
An integral in Eq.~(\ref{eq:sigma22DivNon}) is the only divergent contribution to $\sigma_{22}^2$.

\subsection{Calculation of $\Delta\sigma_{\delta13}^2$}
\label{sec:sigma13}

An expression for $\sigma_{13}^2$, given by Eq.~(\ref{eq:sigmaNonSym13}), involves the third-order density kernel.

\subsubsection{Zel'dovich approximation}
\label{sec:Zeld}

In the Zel'dovich approximation (ZA), the third-order symmetrized density kernel reads
\begin{equation}
F_3^{({\rm ZA})}(\mathbf{k}_1,\mathbf{k}_2,\mathbf{k}_3)=\frac{1}{3!}\,\frac{(\mathbf{k}\cdot\mathbf{k}_1)(\mathbf{k}\cdot\mathbf{k}_2)(\mathbf{k}\cdot\mathbf{k}_3)}{k_1^2k_2^2k_3^2}\,,
\label{eq:F3Zel}
\end{equation}
where $\mathbf{k}=\mathbf{k}_1+\mathbf{k}_2+\mathbf{k}_3$. Hence $F_3^{({\rm ZA})}(\mathbf{k},\mathbf{q},-\mathbf{q})=-(k^2/q^2)\,\mu_{kq}^2/6$ and
\begin{equation}
6F_{3s}^{({\rm ZA})}(\mathbf{k},\mathbf{q})=(1-2a^{-2})\mu_{kq}^2\,.
\label{eq:F3sZel}
\end{equation}
We thus have
\begin{equation}
\begin{split}
\hspace{-.5cm}\sigma_{\delta13}^{2({\rm ZA})}={}&\int\frac{\mathrm{d}k\,\mathrm{d}q}{(2\pi^2)^2}k^2P(k)q^2P(q)\int\frac{\mathrm{d}\Omega_k\,\mathrm{d}\Omega_q}{(4\pi)^2}
\left[\frac{L_0(\mu_{kq})+2L_2(\mu_{kq})}{3}\right]\\
&-\frac{1}{(2\pi)^6}\int\mathrm{d}^3k\,P(k)\int\mathrm{d}^3q\,P(q)\frac{2\mu_{kq}^2}{[a(k,q)]^2}.
\end{split}
\label{eq:sigma13Zel}
\end{equation}
Adding $\sigma_{\delta22}^2$, Eq.~(\ref{eq:sigma22DivNon}), to $\sigma_{\delta13}^2$, Eq.~(\ref{eq:sigma13Zel}), we see that the divergent terms exactly cancel and we obtain
\begin{equation}
\sigma_{\delta13}^{2({\rm ZA})}+\sigma_{\delta22}^{2({\rm ZA})}=2\left(\frac{1}{6}+c_0^2+\frac{2}{3}c_0c_2+\frac{1}{5}c_2^2\right)\sigma_L^4\,
\end{equation}
As mentioned above, in the case of the ZA, $c_0=c_2=1/2$. Therefore [see Eq.~(\ref{eq:sigma-total})] the final result is \citep{Scocci}
\begin{equation}
\sigma_{\delta}^{2({\rm ZA})}=\sigma_L^2+\frac{19}{15}\sigma_L^4\,.
\label{eq:sigmaZAfinal}
\end{equation}

\subsubsection{Exact dynamics}
\label{sub:exact}

\begin{figure}
\centering
\includegraphics[width=\columnwidth]{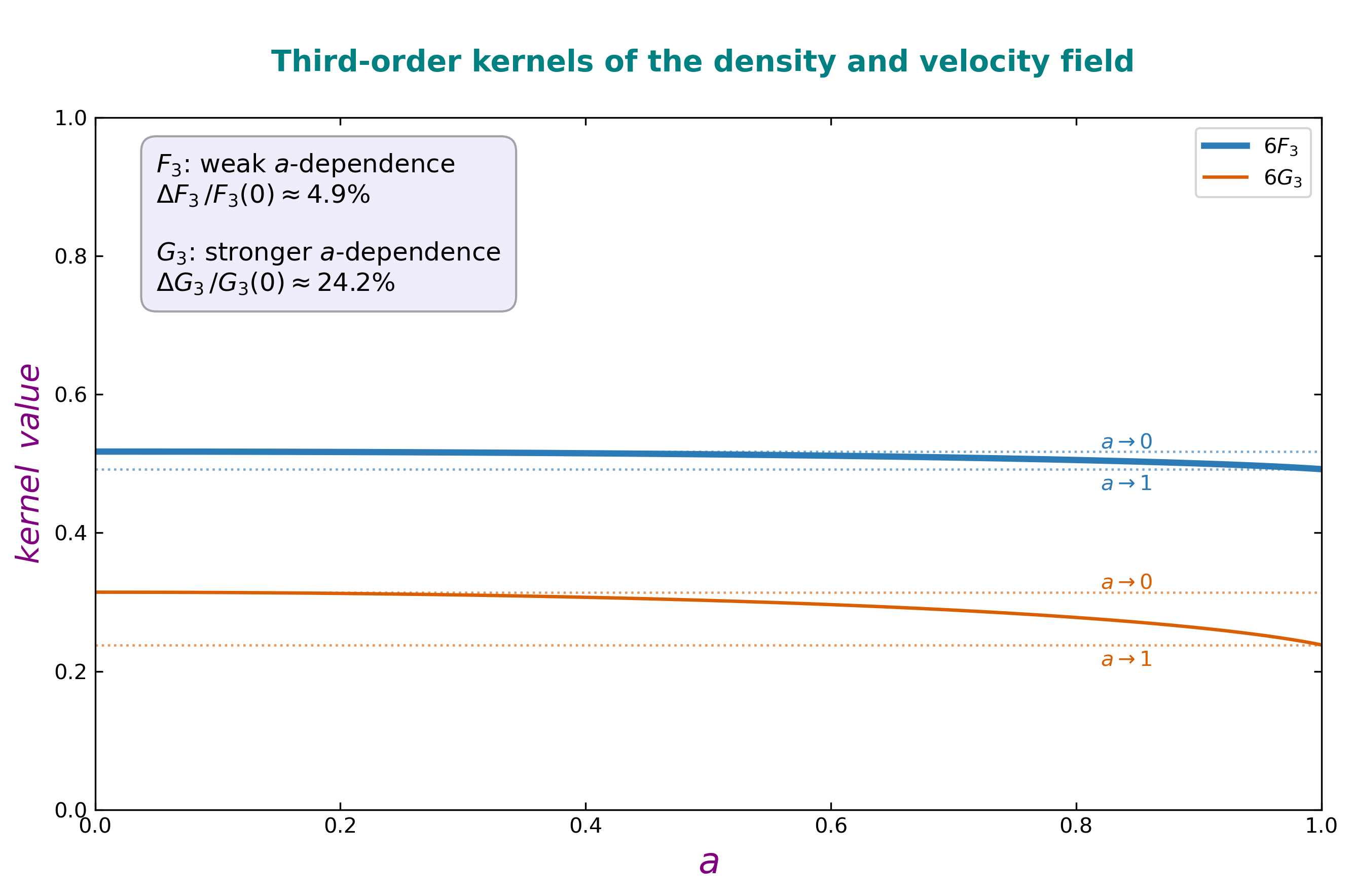}
\caption{Angularly averaged non-divergent contributions of the density and velocity-divergence kernels.}
\label{fig:F3aNonDiv}
\end{figure}

Unlike the expression for $F_3(\mathbf{k},\mathbf{q},-\mathbf{q})$, the expression for $F_{3s}(\mathbf{k},\mathbf{q})$, Eq.~(\ref{eq:F3Sym}), is simple. It can be cast to the following form:
\begin{equation}
6F_{3s}(\mathbf{k},\mathbf{q})=-2\frac{\mu_{kq}^2}{a^2}+6F_{3s}^{({\rm non})}(a,\mu_{kq})\,,
\end{equation}
where the non-divergent part is
\begin{equation}
6F_{3s}^{({\rm non})}=\frac{24}{27}\mu_{kq}^2+\frac{(20a^2-12)\mu_{kq}^4+(4-27a^2)\mu_{kq}^2+15}{63(1-a^2\mu_{kq}^2)}\,.
\end{equation}
We now angularly {\sl average} $6F_{3s}^{({\rm non})}(a,\mu_{kq})$, $6F_{3s}^{({\rm non})}(a)\equiv\int_{-1}^{1}F_{3s}^{({\rm non})}(a,\mu_{kq})\,\mathrm{d}\Omega_k\mathrm{d}\Omega_q/(4\pi)^2$. The result is
\begin{equation}
6F_{3s}^{({\rm non})}(a)=\frac{39a^4-20a^2+12}{63a^4}-\frac{4(1-a^2)^2\tanh^{-1}(a)}{21a^5}\,.
\end{equation}
Figure~\ref{fig:F3aNonDiv} shows $6F_{3s}^{({\rm non})}$ as a function of $a$.

We have
\begin{equation}
\sigma_{\delta13}^2=6\int_0^\infty\frac{\mathrm{d}k\,k^2}{2\pi^2}P(k)\int_0^\infty\frac{\mathrm{d}q\,q^2}{2\pi^2}P(q)F_{3s}^{({\rm non})}(a)\,.
\end{equation}
To proceed further with calculations, a change of variables is necessary. However, note that $F_{3s}^{({\rm non})}(a)$ is remarkably close to a constant. Therefore, without any further effort we can put tight constrains on all possible values of $\sigma_{\delta13}^2/\sigma_L^4$. As $6F_{3s}^{({\rm non})}$ decreases monotonically from $6F_{3s}^{({\rm non})}(0)=4007/2205$ to $6F_{3s}^{({\rm non})}(1)=4063/2205$, we obtain
\begin{equation}
\frac{\sigma_{\delta13}^2}{\sigma_L^4}\in\left[\frac{4007}{2205},\frac{4063}{2205}\right],
\end{equation}
or
\begin{equation}
\sigma_{\delta13}^2/\sigma_L^4=1.830 \pm 0.013\,.
\end{equation}
The result of \citet{Scocci}, $\sigma_{\delta 13}^2\simeq1.82$, is consistent with ours.
\subsubsection{Power-law power spectra}
\label{subsec:power-law}

In this Section we will calculate specific results for the power-law form of the linear power spectra, $P(k)=Ak^n$. In order to make the variance finite, for spectral indices $n>-3$, we introduce an ultraviolet cutoff, $k_c$. We then have
\begin{equation}
\sigma_L^4=\frac{A^2k_c^{2(n+3)}}{8\pi^4(n+3)}\,.
\end{equation}
We change the scalar variables of integration, $k$ and $q$, to $a$ [Eq.~(\ref{eq:a})] and
\begin{equation}
p=(k+q)/2\,.
\end{equation}
After some algebra, the determinant of the Jacobian transformation from $(k,q)$ to $(p,a)$ is
\begin{equation}
|\det J|=\frac{2p}{(1-a)^{1/2}(1+a)^{3/2}}
\end{equation}
and
\begin{equation}
(kq)^{n+2}|\det J|=\frac{2^{n+3}a^{n+2}}{(1-a)^{1/2}(1+a)^{n+7/2}}p^{2n+5}.
\end{equation}
We split the region of integration over $k$ and $q$ into two adjacent triangles,
\begin{equation}
\int_0^{k_c}\mathrm{d}k\int_0^{k_c}\mathrm{d}q=\int_0^{k_c}\mathrm{d}k\int_k^{k_c}\mathrm{d}q+\int_0^{k_c}\mathrm{d}k\int_0^k\mathrm{d}q.
\label{eq:split}
\end{equation}
As both $p$ and $a$ are symmetric functions of $k$ and $q$, the first integral in the RHS of Eq.~(\ref{eq:split}) is equal to the second, where $q/k<1$. 

Then
\begin{equation}
\frac{\pi^4\sigma_{\delta13}^2(n)}{A^2 2^{2(n+1)}}=\int_0^1\frac{a^{n+2}F_{3s}^{({\rm non})}(a)\,\mathrm{d}a}{(1-a)^{1/2}(1+a)^{n+7/2}}\int_0^{f(a)k_c}p^{2n+5}\,\mathrm{d}p,
\end{equation}
where
\begin{equation}
f(a)=\frac{1+a-(1-a^2)^{1/2}}{2a}k_c\,.
\end{equation}
Integrating over $p$ we obtain
\begin{equation}
\frac{\sigma_{\delta13}^2}{\sigma_L^4}=3\cdot2^{2n+5}\int_0^1\frac{a^{n+2}f^{2(n+3)}(a)F_{3s}^{({\rm non})}(a)}{(1-a)^{1/2}(1+a)^{n+7/2}}\,\mathrm{d}a\,.
\end{equation}
Table~\ref{tab:power-law-density} shows non-linear contributions to the density variance in units of $\sigma_L^4$, for power-law power spectra with spectral index $n$.

\begin{table}
\caption{Non-linear contributions to the density variance in units of $\sigma_L^4$, for power-law power spectra with spectral index $n$.}
\label{tab:power-law-density}
\centering
\setlength{\tabcolsep}{4pt}   
\renewcommand{\arraystretch}{1.4} 
\begin{tabular}{c c c c}
\hline\hline
Index & Contribution & Contribution & Total value\\
$n$ & $2F_2^{2({\rm non})}$ & $6F_{3s}^{({\rm non})}$ & $2F_2^{2({\rm non})}+6F_{3s}^{({\rm non})}$\\ [0.4ex]
\hline
$-3$ & $974/735$ & $163/315$ & $1.843$\\ 
$-2$ & $974/735$ & $(1616-9\pi^2)/3024$ & $1.830$\\ 
$-1$ & $974/735$ & $(3839-768\ln2)/6615$ & $1.825$\\ 
$0$  & $974/735$ & $(736+27\pi^2)/2016$ & $1.822$\\ 
$1$  & $974/735$ & $(7709+3072\ln2)/19845$ & $1.821$\\
\hline
\end{tabular}
\end{table}
\subsection{Velocity-divergence field}
\label{sec:vel}

The second-order kernel of the velocity-divergence field is
\begin{equation}
G_2(\mathbf{k},\mathbf{q})=d_0+a^{-1}\mu_{kq}+d_2\mu_{kq}^2,
\label{eq:G_2}
\end{equation}
where $d_0+d_2=1$; specifically
\begin{equation}
d_0=3/7,\quad d_2=4/7\quad ({\rm ED});\qquad d_0=0,\quad d_2=1\quad ({\rm ZA})\,.
\end{equation}

Inspecting Eq.~(\ref{eq:sigma22DivNon}) we see that the non-divergent part of $\sigma_{22}^2$ in the case of the velocity divergence is
\begin{equation}
\frac{\sigma_{\theta22}^2}{\sigma_L^4}=2\left(d_0^2+\frac{2}{3}d_0d_2+\frac{1}{5}d_2^2\right)=\frac{202}{245}
\label{eq:sigma22Vel}
\end{equation}
for exact dynamics and $2/5$ for the Zel'dovich approximation. Using recurrence relations for perturbative kernels in the Zel'dovich approximation [see Eqs.~(2.15a)--(2.16c) of \citet{Scocci}], one can show that
\begin{equation}
G_{3s}^{({ \rm ZA})}(\mathbf{k},\mathbf{q})=F_{3s}^{({\rm ZA})}(\mathbf{k},\mathbf{q})=\frac{1}{6}(1-2a^{-2})\mu_{kq}^2\,.
\label{eq:G3sZel}
\end{equation}
We can thus write immediately
\begin{equation}
\frac{\sigma_{\theta13}^{2({\rm ZA})}+\sigma_{\theta22}^{2({\rm ZA})}}{\sigma_L^4}=2\left(\frac{1}{6}+d_0^2+\frac{2}{3}d_0d_2+\frac{1}{5}d_2^2\right)=\frac{11}{15}\,.
\end{equation}
Note that this coefficient is smaller than that for the density field [Eq.~(\ref{eq:sigmaZAfinal})].

Using exact-dynamics recurrence relations for perturbative kernels, we calculate exact $G_{3s}(\mathbf{k},\mathbf{q})$,
\begin{equation}
6G_{3s}(\mathbf{k},\mathbf{q})=-2\frac{\mu_{kq}^2}{a^2}+6G_{3s}^{({\rm non})}(a,\mu_{kq})\,,
\end{equation}
where the non-divergent part is
\begin{equation}
6G_{3s}^{({\rm non})}(a,\mu_{kq})=\frac{-1+3(4-a^2)\mu_{kq}^2-4(1+a^2)\mu_{kq}^4}{7(1-a^2\mu_{kq}^2)}\,.
\end{equation}
We {\sl average} $6G_{3s}^{({\rm non})}(a,\mu_{kq})$ over all values of $\mu_{kq}$, $6G_{3s}^{({\rm non})}(a)\equiv 3\int_{-1}^1G_{3s}^{({\rm non})}(a,\mu_{kq})\,\mathrm{d}\mu_{kq}$. The result is
\begin{equation}
6G_{3s}^{({\rm non})}(a)=\frac{13a^4-20a^2+12}{21a^4}-\frac{4(1-a^2)^2\tanh^{-1}(a)}{7a^5}\,.
\end{equation}

Surprisingly, $G_{3s}^{({\rm non})}(a)$ is a linear function of $F_{3s}^{({\rm non})}(a)$. More specifically,
\begin{equation}
6G_{3s}^{({\rm non})}(a)=3\left[6F_{3s}^{({\rm non})}(a)\right]-\frac{26}{21}\,.
\end{equation}
We can thus write in general
\begin{equation}
\frac{\sigma_{\theta13}^2}{\sigma_L^4}=3\frac{\sigma_{\delta13}^2}{\sigma_L^4}-\frac{26}{21}\,.
\end{equation}
Without any additional effort, therefore, we can construct a table which shows non-linear contributions to the variance of the velocity divergence in units of $\sigma_L^4$, for power-law power spectra with spectral index $n$. It is given by Table~\ref{tab:power-law-velocity}.

\section{Conclusions}
\hspace{.5cm}We have calculated the leading-order corrections to the variances of the cosmic density and velocity fields using both the exact dynamics and the Zel'dovich approximation. For the power-law forms of the power spectra, our results are in exact agreement with those obtained by \citet{Scocci}. Beyond this special case, our novel approach enabled us to put robust constraints on the nonlinear variance of the density field for {\em any} form of the power spectrum. We found that the corresponding coefficient of the leading-order correction is confined to a remarkably narrow interval of 4007/2205 to 4063/2205 $(\simeq 1.817 \leq C \leq 1.843)$, quite independently of the shape of the initial power spectrum.
\begin{table}
\caption{Non-linear contributions to the velocity variance in units of $\sigma_L^4$, for power-law power spectra with spectral index $n$.}
\label{tab:power-law-velocity}
\centering
\setlength{\tabcolsep}{2.9pt}
\renewcommand{\arraystretch}{1.4}
\begin{tabular}{c c c c}
\hline\hline 
Index & Contribution & Contribution & Total value\\ 
$n$ & $2G_2^{2({\rm non})}$ & $6G_{3s}^{({\rm non})}$ & $2G_2^{2({\rm non})}+6G_{3s}^{({\rm non})}$\\ [.4ex]
\hline 
$-3$ & $202/245$ & $11/35$ & $1.139$\\ 
$-2$ & $202/245$ & $(368-9\pi^2)/1008$ & $1.101$\\ 
$-1$ & $202/245$ & $(1109-768\ln2)/2205$ & $1.086$\\ 
$0$  & $202/245$ & $(9\pi^2-32)/224$ & $1.078$\\ 
$1$  & $202/245$ & $(3072\ln2-481)/6615$ & $1.074$\\
\hline
\end{tabular}
\end{table}

\nocite{*}
\bibliographystyle{aa}
\bibliography{references}


\end{document}